\documentclass[11pt]{article}
\usepackage{amssymb}
\usepackage{epsfig}

\newcommand{\dl}{\delta^{(3)}({\bf r})}

\newcommand{\BE}{\begin{equation}}
\newcommand{\EE}{\end{equation}}
\newcommand{\BA}{\begin{eqnarray}}
\newcommand{\EA}{\end{eqnarray}}

\begin{document}
\begin{titlepage}
\begin{center}

            {\LARGE{\bf Ultraweak excitations of the quantum vacuum
            \\ as physical models of gravity }}

\vspace*{14mm} {\Large M. Consoli }
\vspace*{4mm}\\
{\large
Istituto Nazionale di Fisica Nucleare, Sezione di Catania \\
Via Santa Sofia 64, 95123 Catania, Italy}

\end{center}
\begin{center}
{\bf Abstract}
\end{center}

It has been argued by several authors that the space-time curvature
observed in gravitational fields, and the same idea of forms of
physical equivalence different from the Lorentz group, might emerge
from the dynamical properties of the physical flat-space vacuum in a
suitable hydrodynamic limit. To explore this idea, one could start
by representing the physical vacuum as a Bose condensate of
elementary quanta and look for vacuum excitations that, on a coarse
grained scale, resemble the Newtonian potential. In this way, it is
relatively easy to match the weak-field limit of classical General
Relativity or of some of its possible variants. The idea that Bose
condensates can provide various forms of gravitational dynamics is
not new. Here, I want to emphasize some genuine quantum field
theoretical aspects that can help to understand i) why
infinitesimally weak, $1/r$  interactions can indeed arise from the
same physical vacuum of electroweak and strong interactions and ii)
why, on a coarse-grained scale, their dynamical effects can be
re-absorbed into an effective curved metric structure.

\vskip 35 pt
\end{titlepage}

\section{Introduction}

The usual interpretation of phenomena in gravitational fields is in
terms of a fundamentally curved space-time. However, showing the
consistency of this interpretation with some basic aspects of the
quantum theory has been proven to be a very difficult problem. For
this reason, one could try to explore a completely different
approach where an {\it effective } curvature reflects
long-wavelength distortions of the same physical, flat-space vacuum
(compare e.g. with the curvature of light in Euclidean space when
propagating in a medium with variable density).

Looking at gravity in this perspective, it is useful to start by
first exploring those systems (moving fluids, condensed matter
systems with a refractive index, Bose-Einstein condensates,...) that
can simulate the effects of a genuine space-time curvature. For
these systems, at a fundamental level, space-time is exactly flat.
However, an effective curved metric emerges
 when describing the propagation of low-energy
fluctuations. Thus, this `emergent-gravity' approach \cite{barcelo}
explores the possibility that the type of description of classical
General Relativity (and of its possible variants) might be similar
to hydrodynamics that, concentrating on the properties of matter at
scales larger than the mean free path for the elementary
constituents, is insensitive to the details of the underlying
molecular dynamics. In some way, the same type of idea was also at
the basis of the original `induced-gravity' approach \cite{adler}.

For a definite and simple example where one can clearly separate out
the various aspects of the problem, let us consider Visser's analogy
\cite{visser} with a moving irrotational fluid , i.e. where the
velocity field is the gradient of a scalar potential $s (x)$. In
this system, the underlying space-time is exactly flat but the
propagation of long-wavelength fluctuations is governed by a curved
effective metric $g_{\mu\nu}(x)$ determined at each space-time point
$x$ by the physical parameters of the fluid (density, velocity and
pressure). These can all be expressed in terms of $s(x)$, and of its
derivatives, through the hydrodynamical equations. In this example,
there is a system, the fluid, whose elementary constituents are
governed by some underlying molecular interactions that, on some
scale, can be summarized in the value of a scalar field $s(x)$. At
some intermediate level, $s(x)$ contains the relevant dynamical
informations. On the other hand, the effective metric tensor
$g_{\mu\nu}$ is a derived quantity that depends on $x$ in some
parametric form \BE \label{parametric}
         g_{\mu \nu}(x)=g_{\mu \nu}[s(x)]
\EE The fluid analogy is also interesting for another reason.
According to present particle physics, the physical vacuum is not
trivially empty but is filled by particle condensates that play a
crucial role in many fundamental phenomena such as particle mass
generation and quark confinement. In the physically relevant case of
the standard model of electroweak interactions, this can be
summarized by saying \cite{thooft} that "What we experience as empty
space is nothing but the configuration of the Higgs field that has
the lowest possible energy. If we move from field jargon to particle
jargon, this means that empty space is actually filled with Higgs
particles. They have Bose condensed". Thus, it becomes natural to
represent the vacuum as a superfluid medium, a quantum liquid.

As pointed out by Volovik \cite{volo}, in this representation, if
the inducing-gravity scalar field $s(x)$ were identified with some
excitation of such a vacuum state, i.e. with a function that
vanishes exactly in the unperturbed state, it would be easy to
understand why there is no non-trivial curvature in the equilibrium
state $s(x)=0$ where any liquid is self-sustaining. Namely, in the
ground state, space-time would look exactly flat \BE \label{flat}
         g_{\mu \nu}[s=0]=\eta_{\mu \nu}={\rm
         diag}(1,-1,-1,-1)
\EE just because the large condensation energy of the medium plays
no role. In any liquid, in fact, curvature requires {\it deviations}
from the equilibrium state. The same happens for a crystal at zero
temperature where all lattice distortions vanish and electrons can
propagate freely as in a perfect vacuum. In both cases, the large
condensation energies of the liquid and of the crystal do not
generate any curvature. This is the most dramatic difference from
the standard approach where, in principle, all forms of energy
gravitate and generate a curvature. This point of view represents
the simplest and most intuitive solution of the so called
cosmological-constant problem found in connection with the energy of
the quantum vacuum. In this sense, by exploring emergent-gravity
approaches based on an underlying superfluid medium, one is also
taking seriously Feynman's indication : "...the first thing we
should understand is how to formulate gravity so that it doesn't
interact with the energy in the vacuum" \cite{rule}.

However, before starting with any analysis of the physical vacuum
and of its excitations, other general observations are still
required. A first observation is that, if really gravity were a
long-wavelength modification of the physical condensed vacuum,
explaining its characteristic features might require to use all
possible informations. These include basic elements of the quantum
theory, such as particle-wave duality, or other technical details of
particle physics as with the origin of hadronic masses from the
gluon and chiral condensates, the `triviality' of contact point-like
interactions in 3+1 dimensions and so on. In this way, by putting
together all available theoretical and experimental knowledge,
aspects that are apparently unrelated can finally provide a single
consistent framework.

A second observation is that, due to the complexity of the problem,
it is unlike to get at once a satisfactory description of all
possible phenomena. Thus, even though a scalar field cannot account
for all possible gravitational phenomena, one could, nevertheless,
start with this model, as when describing the longitudinal density
fluctuations of a medium. If, as it happens with many physical
systems (elastic media \cite{sommerfeld}, turbulent fluids
\cite{troshkin,tsankov}, superfluids \cite{wilks},...), one has also
to describe transverse waves, at some later stage one can introduce
a genuine vector field ${\bf V}(x)$ (with $\nabla\cdot{\bf V}=0$)
and replace the parametric dependence of the metric tensor with the
more general structure \footnote{For a definite realization of this
idea see Puthoff's \cite{puthoff2} derivation of the
`gravitomagnetic' and `gravitoelectric' fields, entering linearized
general relativity, from the truncated hydrodynamical equations for
a slightly compressible turbulent fluid. Puthoff's Eq.(41) for the
effective metric has exactly the same form as in
Eq.(\ref{general}).} \BE \label{general} {g}_{\mu \nu}(x)={g}_{\mu
\nu}[s(x),{\bf V}(x)] \EE This type of extension, by itself, would
not pose particular conceptual problems.

A third observation is that, in the presence of a condensed vacuum,
one is tacitly adopting a 'Lorentzian' perspective \cite{sonego},
namely where physical rods and clocks are held together by the same
basic forces entering the structure of the underlying `ether' (the
physical vacuum). Thus the principle of relativity means that the
measuring devices of moving observers are dynamically affected in
such a way that their uniform motions with respect to the ether
frame become undetectable. In this sense, one is naturally driven to
consider the possible, coarse-grained forms of effective curved
metric structures as originating from a re-definition of the basic
space-time units.

This aspect was well summarized by Atkinson as follows
\cite{atkinson} : "It is possible, on the one hand, to postulate
that the velocity of light is a universal constant, to define {\it
natural} clocks and measuring rods as the standards by which space
and time are to be judged and then to discover from measurement that
space-time is {\it really} non-Euclidean. Alternatively, one can
{\it define} space as Euclidean and time as the same everywhere, and
discover (from exactly the same measurements) how the velocity of
light and natural clocks, rods and particle inertias {\it really}
behave in the neighborhood of large masses." Such a type of
correspondence, in fact, is known to represent one of the  possible
ways to introduce the concept of curvature, see e.g. \cite{feybook}.

By adopting this point of view, one could start by assuming that,
under the influence of $s(x)$, any mass scale $M$ (or binding
energy) might be replaced by an effective mass, say \BE
\label{rescaling1} M \to {{M}\over{\lambda(s)}} \EE At the same
time, if the scalar function $s(x)$ were somehow describing the
density fluctuations of the vacuum medium, it would be natural to
introduce, besides such a re-scaling, a vacuum refractive index
${\cal N}(s)$. By considering the vacuum as a `non-dispersive'
transparent medium, this could take into account the different
geometrical constraints that are placed, on the wavelength of light
of a given frequency, by the presence of these fluctuations.
Equivalently, it could be due to the re-definition of the local
vacuum energy and, with it, of the local dielectric constants as in
the polarizable-vacuum approach of
Refs.\cite{wilson,dicke,varenna,puthoff}. In any case, one is driven
to consider the effective metric structure \BE \label{basic01}
         {g}_{\mu \nu}[s]\equiv
{\rm diag}({{\lambda^2(s) }\over{{\cal
N}^2(s)}},-\lambda^2(s),-\lambda^2(s),-\lambda^2(s)) \EE that
re-absorbs the local, isotropic modifications of space-time into its
basic ingredients: the value of the speed of light and the
space-time units.

It is interesting that, independently of the specific underlying
mechanisms that one can imagine to generate ${\cal N}(s)$ and
$\lambda(s)$, these two functions can further be related through
general arguments that express the basic property of light of being,
at the same time, a corpuscular and undulatory phenomenon. In flat
space, this particle-wave `duality' reflects the equivalence of the
speed of light defined as a `particle' velocity from the condition
$ds^2=0$ with that obtained from the solutions of the D'Alembert
wave equation $\Box F=0$.

To consider the analogous situation in curved space, let us assume
$s=0$ at infinity (where $\lambda={\cal N}=1$) and consider a
solution of the wave equation that describes asymptotically a
monochromatic signal of definite frequency $\omega$ and wave vector
$k$. By re-writing Eq.(\ref{basic01}) as a general isotropic metric
\BE \label{isotropic}
 {g}_{\mu \nu} \equiv {\rm diag}(A,-B,-B,-B)
\EE one may ask under which conditions the local speed of light
$\sqrt{ {{ A}\over{B}} }$, defined from the condition
$ds^2={g}_{\mu\nu} dx^\mu dx^\nu=0$, agrees with that obtained from
the covariant D'Alembert wave equation \cite{progress} \BE
\label{dalembert} {{1}\over{A}} {{\partial^2 F }\over{\partial t^2}}
-{{1}\over{B}}( {{\partial^2}\over{\partial x^2}}+
{{\partial^2}\over{\partial y^2}}+ {{\partial^2}\over{\partial
z^2}})F - {{1}\over{\sqrt {AB^3} }} (\nabla \sqrt{AB}) \cdot (\nabla
F)=0 \EE or, by introducing the 3-vector ${\bf{g}}\equiv \sqrt {
{{A}\over {B^3}} } (\nabla \sqrt{AB})$, \BE
\label{omega}{{1}\over{F}} {{\partial^2 F }\over{\partial t^2}}
              = {{A}\over{B}}~ {{1}\over{F}}\Delta F + {{1}\over{F}}
              {\bf{g}}\cdot (\nabla F)
\EE Thus, by identifying ${{1}\over{F}}{{\partial^2 F
}\over{\partial t^2}}$ as the local equivalent of $-\omega^2$ and
${{1}\over{F}}\Delta F$ as the corresponding one for $-k^2$, one
finds that particle velocity and the `phase velocity'
${{\omega}\over{k}}$ agree with each other only when ${\bf{g}}=0$,
i.e. when $AB$ is a constant. This product can be fixed to unity
with flat-space boundary conditions at infinity and, therefore, the
resulting value \BE AB=1 \EE or, in our case \BE \label{index} {\cal
N}(s)=\lambda^2(s) \EE can be considered a consistency requirement
on the possible modifications of the underlying physical vacuum, if
these modifications have to preserve, at least to some definite
order in powers of $s(x)$, the observed particle-wave duality which
is intrinsic in the nature of light.

In a more technical language, one could say that this type of
particle-wave duality, among all possible forms of the covariant
D'Alembert wave equation
($g^{\alpha\beta}g_{\beta\gamma}=\delta^\alpha_\gamma$) \BE \Box_c
F=g^{\alpha\beta} {{
\partial^2 F } \over{\partial x^\alpha
\partial x^\beta }} - \Gamma^{\nu} {{\partial F}\over{\partial
x^{\nu} }}=0 \EE selects the harmonic-coordinate condition \BE
\label{harmonic}  \Gamma^\nu=
g^{\alpha\beta}\Gamma^\nu_{\alpha\beta}=0 \EE More in general, the
special role of harmonic coordinates, when imposing flat-space
boundary conditions at infinity, had been strongly emphasized by
Fock \cite{fock}. In his view, one should not confuse the general
covariance of a set of partial differential equations with the
notion of physical equivalence that is used to formulate a principle
of relativity. The former is a logical requirement in all cases
where the coordinate system is not fixed in advance. The latter, on
the other hand, is related to the existence of frames of reference
of a certain class for which one can define corresponding physical
processes and, in the presence of a given gravitational field,
depends on the specific boundary conditions. These boundary
conditions cannot be given in a general-covariant form so that the
fact that for determinateness Einstein's ten equations must be
supplemented by four additional conditions is in itself evident. For
physical equivalence, what really matters is not the possibility of
an indeterminate formulation but just the opposite, namely a
formulation that is as determinate as is allowed by the nature of
the problem as, for instance, by the requirement of uniformity at
infinity. In this sense, Fock concludes that the harmonic
coordinates could be considered the analogue of the inertial systems
when dealing with gravitational fields that vanish asymptotically.
As such, it is not surprising that they may enjoy other special
properties (e.g. particle-wave duality) that do not hold in other
arbitrary coordinate systems. Here I want to emphasize that, in an
emergent-gravity approach, the selection of this class of
coordinates might be a natural consequence of the vacuum structure.

To consider in some more detail the issue of general covariance in
this type of approach, let us start from Einstein's original idea,
namely to consider forms of physical equivalence that could be
naturally  described within a general-covariant formalism
\footnote{"...the set of all transformations in any case includes
those which correspond to all relative motions of three-dimensional
systems of coordinates" \cite{dover}.}. To understand how these
forms of physical equivalence could originate from the hydrodynamic
limit of the same flat-space vacuum, let us tentatively assume that,
on a coarse-grained scale, $s(x)$ resembles the Newtonian potential.
Then, particle trajectories in this field would not depend on the
particle mass thus providing a basic ingredient to represent
dynamical effects as an overall modification of the space-time
geometry. At the same time, once $s(x)$ were coupling universally to
the various forms of matter, there would be the possibility of
establishing an analogy between the motion of a body in a
gravitational field and the motion of a body, not subject to an
external field, but viewed by a non-inertial reference frame.
Therefore, it is this relation with the non inertial forces that
produces new forms of physical equivalence (i.e. different from the
simplest uniform translational motions) and leads naturally to adopt
a general-covariant formulation. This type of derivation could even
be pursued to argue that the parametric dependence of the effective
metric tensor ${g}_{\mu \nu}(x)={g}_{\mu \nu}[s(x)]$ should
correspond to solve Einstein's field equations with a suitable
stress tensor that, in addition to the standard matter
contributions, might also depend on the $s-$ field. In this way, one
could partially fill the conceptual gap with classical General
Relativity or with some of its possible variants.

For a definite realization of this idea, let us consider Yilmaz'
original approach \cite{yilmaz}. In his view, the Newtonian
potential that solves the Poisson equation in flat space for a given
mass distribution  (I set $c=1$ and denote by $G_N$ the Newton
constant) \BE\label{newton} s(x)= -G_N \sum_k {{M_k}\over{|{\bf x} -
{\bf x}_k|}} \EE is the true agent of gravity. On the other hand,
the metric tensor
 \BE \label{yilmaz} {g}_{\mu\nu}[s]={\rm diag}(e^{2s},-
e^{-2s},- e^{-2s},- e^{-2s}) \EE that solves Einstein's field
equations, with a stress tensor for the $s-$ field $t^\mu_\nu(s)=
-\partial^\mu s\partial_\nu s + 1/2\delta^\mu_\nu~\partial^\alpha
s\partial_\alpha s$, is a derived quantity that depends on $s(x)$ in
a parametric form \footnote{In this sense, Yilmaz' original
formulation could be considered the prototype of all
emergent-gravity approaches based on a parametric dependence of the
effective metric tensor on (scalar, vector, tensor,...) excitations
of the flat-space vacuum. In the long-wavelength limit, these
excitations determine self-consistently the effective metric through
their contributions to the energy-momentum tensor. }. In spite of
the obvious differences, in the one-body case and in the weak-field
limit, one can expand the Schwarzschild metric of General Relativity
as \BE g_{00}=\left( {{ 1 +s/2 }\over{ 1 -s/2 }} \right)^2= 1+2s +
2s^2+..\sim e^{+2s} \EE \BE g_{11}=g_{22}=g_{33}=-\left( 1
-s/2\right)^4 = -(1-2s+..)\sim -e^{-2s} \EE and get the same
agreement with experiments to the present level of accuracy
\footnote{For a discussion of this point, in the original Yilmaz
theory, see Ref.\cite{tupper}. Cosmological implications, in more
recent formulations of the Yilmaz approach, are also discussed in
Ref.\cite{ibison}}.

Thus, to match both metric structures, it should be possible to show
i) that scalar excitations that resemble the Newtonian potential
might indeed arise from the same physical vacuum of present particle
physics ii) that under their influence one can expand the re-scaling
of Eq.(\ref{rescaling1}) as \BE \label{rescaling2}
{{1}\over{\lambda(s)}}= 1+s +{\cal O}(s^2) \EE This leads to the
line element \BE \label{basic02}
         {g}_{\mu \nu}[s]\equiv
{\rm diag}((1+2 s) ,-(1-2s),-(1-2s),-(1-2s)) \EE that agrees with
the first approximation both in General Relativity and in the Yilmaz
approach.

The idea that Bose condensates can provide various forms of
gravitational dynamics is not new (see e.g.
\cite{bosegravity,sindoni} and references quoted therein). In the
following four sections I will illustrate the nature of the
hydrodynamic limit and discuss some genuine quantum field
theoretical aspects that could be useful to understand how an
infinitesimally weak, long-range interaction as the Newtonian
potential, could indeed arise from the same physical vacuum of
electroweak and strong interactions.

It is interesting, however, that in principle, in some extreme
situations, these ultraweak effects could become important. In fact,
the ambiguity about the higher-order ${\cal O}(s^2)$ effects means
that, for large massive bodies, the energy content of the $s-$field
could become so strong to screen the Schwarzschild singularity
expected for a single point-like mass. These issues will be briefly
addressed in the final section, together with a more general
discussion of the virtues and present limitations of the approach.

\section{Excitations of the scalar condensate}

As anticipated, the fundamental phenomenon of symmetry breaking,
that is believed to determine the physical vacuum of electroweak
interactions, consists in the spontaneous creation from the empty
vacuum of elementary spinless quanta and in their macroscopic
occupation of the same quantum state. The translation from `field
jargon to particle jargon'  can be obtained, for instance, along the
lines of Ref.\cite{mech}. This amounts to establish a well defined
functional relation $n=n(\phi^2)$ between the average density $n$ of
scalar quanta in the ${\bf{k}}=0$ mode and the average value $\phi$
of the scalar field. Thus, Bose condensation is just a consequence
of the shape of the effective potential of the theory whose
stability requires values of $\phi$ such that $n\neq 0$.

In this framework, a simple unified picture of the underlying scalar
system can be given in terms of two basic quantities, the particle
density ${n}$ of the elementary condensing quanta and their
scattering length $a$ that represents the quantum-mechanical
analogue of the hard-sphere radius introduced in a classical
description. In terms of these quantities, one finds the order of
magnitude estimate \cite{mech} \BE
                     m_h \sim \sqrt {{n}a}
\EE $m_h$ being the parameter associated with the massive
excitations of the condensed vacuum, usually denoted as the massive
Higgs boson. In this representation, the two quantities ${n}$ and
$a$ can be combined to form a {\it hierarchy} of length scales \BE a
\ll {{1}\over{ \sqrt{ {n}a} }} \ll {{1}\over{ {n}a^2}} \EE that
decouple for an infinitely dilute system  where \BE \label{epsilon}
{n}a^3 \to 0 \EE This hierarchical situation is obtained when
approaching the continuum limit of quantum field theory where, due
to the basic `triviality' property of the underlying contact
$\Phi^4$ theory in 3+1 space-time dimensions \cite{triviality}, the
scattering length $a$ should vanish in units of the typical
elementary particle length scale  \BE \xi_h=1/m_h\sim {{1}\over{
\sqrt{ {n}a} }} \EE Thus, the `triviality' limit, where $a m_h \to
0$, can be simulated by an ultraviolet cutoff \BE \Lambda \sim 1/a
\EE such that $m^2_h/\Lambda^2 \sim {n}a^3 \to 0 $ \footnote{Notice
that the average inter-particle distance $d\sim n^{-1/3}$, although
much larger than the scattering length $a$, is also much smaller
than the length scale $\xi_h$. This means that the scalar condensate
could be considered as infinitely dilute or as infinitely dense
depending on the adopted unit of length. This type of situation is
characteristic of a hierarchical system.}. In the same limit, the
mean free path for the elementary condensed quanta \BE \label{mfp}
r_{\rm mfp} \sim {{1}\over{ {n}a^2}} \EE diverges in units of
$\xi_h=1/\sqrt{ {n}a}$. For instance by choosing $a\sim 10^{-33}$ cm
and a typical electroweak scale $1/m_h=\xi_h \sim 10^{-17}$ cm, one
obtains $r_{\rm mfp} \sim 10^{-1}$ cm. In this way the hydrodynamic
limit of the system, expected for wavelengths larger than $r_{\rm
mfp}$ decouples from the scale $\xi_h$.

Equivalently the region in momentum space $|{\bf{k}}| < \delta$
where \BE \label{delta} \delta\sim 1/r_{\rm mfp} \sim
{{m^2_h}\over{\Lambda}}~, \EE  vanishes in the continuum limit
$\Lambda \to \infty$. In this sense, the hydrodynamic region could
be considered one of the "reentrant violations of special relativity
in the low-energy corner" mentioned by Volovik \cite{volo2}. These
characterize the condensed phase of quantum field theories and can
be used to obtain infinitesimally weak interactions from those
regions of the spectrum that become a zero-measure set in the
continuum limit of the theory.

Now, to construct a model of the Newtonian potential in a Bose
condensate of spinless quanta, one could start from the long-range,
attractive $1/r$ potential derived by Ferrer and Grifols
\cite{ferrer}. This is similar to the long-range attractive
potential among electrons moving inside an ion lattice and can be
qualitatively understood in terms of {\it phonons}, the quantized
long wavelength oscillations of the condensate. Their result can be
re-phrased as follows. Let us consider a scalar field $\Phi(x)$,
whose elementary quanta have some mass $m$, that interacts with an
external density $\rho(x)$ through the Lagrangian \BE {\cal L}_{\rm
int}=  g \rho(x)\Phi^2(x) \EE Our problem is to evaluate the
interaction energy of two density distributions, say $\rho_1({\bf
r})$ and $\rho_2({\bf r})$, centered respectively around ${\bf r}_1$
and ${\bf r}_2$, in the limit of large spatial separations $|{\bf
r}-{\bf r}'|\sim |{\bf r}_1-{\bf r}_2| \to \infty$. This interaction
energy is (minus) the Fourier transform of the Feynman graph with a
loop formed by two $\Phi$ quanta that are exchanged between the
sources. In the trivial empty vacuum of the $\Phi$ field, i.e. when
$\langle \Phi\rangle=0$, this takes the form \BE V_{\rm int}=
-g^2\int{{\rho_1({\bf r})\rho_2({\bf r}') e^{-2m|{\bf r}-{\bf
r}'|}}\over{|{\bf r}-{\bf r}'|^3}}d^3{\bf r}d^3{\bf r}' \EE However,
if the $\Phi$ quanta were Bose condensing below some critical
temperature $T_c$, i.e. when now the vacuum at T=0 has $\langle
\Phi\rangle \neq 0$, the same type of computation gives a very
different interaction, namely \BE V_{\rm int}= -g^2T^2_c
\int{{\rho_1({\bf r})\rho_2({\bf r}')}\over{|{\bf r}-{\bf
r}'|}}d^3{\bf r}d^3{\bf r}' \EE Ferrer and Grifols explain that this
happens because, when one of the two exchanged $\Phi$ quanta is in
the condensate, the other behaves as a massless scalar photon, the
responsible of the Coulomb interaction. This means that the
long-range $1/r$ interaction can be obtained by the replacement \BE
\Phi^2(x) = (\langle \Phi\rangle\ + \delta\Phi(x))^2=\langle
\Phi\rangle^2 + 2\langle \Phi\rangle\delta\Phi(x) + \delta\Phi^2(x)
\EE and picking up the crossed term $2\langle
\Phi\rangle\delta\Phi(x)$. In the presence of a non-zero $\langle
\Phi\rangle$, the propagator of the fluctuation $\delta\Phi(x)$ (the
{\it connected} propagator) behaves as $1/p^2$ for $p_\mu \to 0$.

In general, the connected  propagator $G(x-y)$ determines the
fluctuation of the scalar field $\delta\Phi(x)$, around the average
value $\langle \Phi \rangle$, that is produced by an external
perturbation $J(x)$ through the relation \BE \delta\Phi(x)=\int d^4
x'~ G(x-x')J(x') \EE Thus, for a pointlike, static source, say
$J(x)=J{\delta^{(3)}({\bf r})}$, one finds \BE \delta\Phi({\bf r}
)=J D({\bf r}) \EE with \BE D({\bf r})= \int {{d^3{\bf
p}}\over{(2\pi)^3}}~
 e^{i {\bf p}\cdot{\bf r}}~  G({\bf p},p_4=0) \EE in terms of
the Fourier transform $G(p)$ in Euclidean space \BE G(x)=\int {{d^4{
p}}\over{(2\pi)^4}}~e^{i {p}{ x}}~ G(p)\EE where $p_\mu\equiv({\bf
p},p_4)$. Therefore, if one requires long-range modifications such
that $\delta\Phi$ remains sizeable at large distances from a
perturbing source, the propagator has to become singular when $p_\mu
\to 0$. For instance, in the case of a free massless scalar field,
where $G({\bf p},p_4=0)=1/{\bf p}^2$, one obtains a $\delta\Phi$
that vanishes as $1/r$. On the other hand, in the case of a free
massive scalar field where $G({\bf p},p_4=0)=1/ ({\bf p}^2 +
m^2_h)$, one finds the corresponding short-range Yukawa behaviour
$e^{-m_h r}/r$.

Thus the possibility of a long-range $1/r$ potential in the
broken-symmetry vacuum depends crucially on the zero-momentum limit
of the connected scalar propagator. At the same time, for any given
strength of the source, the magnitude of $\delta \Phi$ depends on
the slope of $G(p)$. The standard unit normalization corresponds to
genuine massless particles (e.g. photons) whose propagator behaves
as $1/p^2$ in the whole range of momenta. However, a condensed
vacuum can exhibit different types of excitations in different range
of momenta. For instance, when $p_\mu \to 0$, the propagator could
behave as $\zeta/p^2$, where $\zeta$ is some positive number, and
approach the massive form at larger $|p|$.

Again, the motivations for this idea originate from the
representation of the broken-symmetry vacuum as a physical
condensate and from the analogy with superfluid $^4$He, the physical
system that is usually considered as a non-relativistic realization
of $\Phi^4$ theory. In fact, as originally proposed by Landau
\cite{landau}, a superfluid should have two different energy
branches, namely gapless density oscillations (phonons) and massive
vortical excitations (rotons)\footnote{The analogy is consistent
with the results of Ref.\cite{hydro} where, by using quantum
hydrodynamics, the mass parameter $m_h\sim \sqrt{na}$ was shown to
be proportional to the energy gap for vortex formation in a suitable
superfluid medium possessing the same constituents and the same
density as in the condensate picture of $\Phi^4$.} . Experiments
however have shown that these two different branches actually merge
into a single energy spectrum, a sort of `hybrid' that smoothly
interpolates between the two different functional forms.

Analogously, in our case, the long-range components of
$\delta\Phi(x)$, for brevity denoted as $s(x)$, would totally be
determined by the infrared part of the propagator. As such, their
interactions, proportional to the $\zeta$ parameter, could be very
different from the typical interaction strength of the
short-wavelength fluctuations. The discussion of these other aspects
will be presented in the following two sections.

\section{The zero-momentum propagator in the broken phase}

The long-range forces considered in the previous section originate
from the macroscopic occupation of the same quantum state. As such,
they are quite unrelated to the Goldstone phenomenon, that
characterizes the spontaneous breaking of continuous symmetries, and
are solely determined by the condensing scalar field, i.e. from that
field $\Phi$ which acquires a non-zero vacuum expectation value. For
this reason, these long-range forces would also exist if spontaneous
symmetry breaking were induced by a one-component field. Their
ultimate origin has to be traced back to a peculiarity of the
zero-momentum connected propagator for the $\Phi-$field: this is a
{\it two-valued} function \cite{consoli}. Namely, its inverse, in
addition to the standard massive solution $G^{-1}_a(p=0)=m^2_h$,
includes the value $G^{-1}_b(p=0)=0$ as in a massless theory. To
show this, one can use different arguments.

Let us introduce preliminarily the semi-classical non-convex
effective potential $V_{\rm NC}(\phi)=V_{\rm NC}(-\phi)$
(NC=Non-Convex), as computed in the standard loop expansion. Let us
also denote by $\phi = \pm v$ its absolute minima and by
$m^2_h\equiv V''_{\rm NC}(\pm v)>0$ its quadratic shape at these
minima. In full generality, one could first study the theory at an
arbitrary value of $\phi$ and then take the $\phi \to \pm v$ limit
afterward. In this case, for any $\phi \neq \pm v$, the diagrammatic
representation of the connected propagator \cite{consoli} requires
to first include the one-particle reducible tadpole graphs where
zero-momentum propagator lines are attached to the one-point
function $\Gamma_1(p=0)=V'_{\rm NC}(\phi)$. By implicitly assuming
the regularity of the zero-momentum propagator, these graphs are
usually ignored {\it at} $\phi=\pm v$ where $V'_{\rm NC}(\pm v)=0$.
Thus, $G^{-1}(p)$ is identified with the 1PI two-point function
$\Gamma_2(p)$, whose zero-momentum value $\Gamma_2(p=0)$ is nothing
but $V''_{\rm NC}(\pm v)$, a positive-definite quantity. On the
other hand, by allowing for a singular $G(p=0)$ at $\phi=\pm v$, one
is faced with a completely different diagrammatic expansion and thus
the simple picture of the broken phase as a pure massive theory,
based on the chain \BE \label{chain}
G^{-1}(p=0)=\Gamma_2(p=0)=V''_{\rm NC}(\pm v)=m^2_h
> 0\EE breaks down.

The existence of two solutions for $G^{-1}(p=0)$ can also be derived
by evaluating in the saddle point approximation the generating
functional $W[J]$ for a constant source and taking the double limit
where $J \to \pm 0$ and the space-time volume $\Omega \to \infty$
\cite{consoli}. As such, the two solutions admit a geometrical
interpretation in terms of left and right second derivatives of the
Legendre-transformed, quantum effective potential. This is convex
downward and is not an infinitely differentiable function when
$\Omega \to \infty$ \cite{syma}. For the convenience of the reader,
this latter type of derivation will be reported below with several
details that were not included in Ref.\cite{consoli}.

To describe spontaneous symmetry breaking in full generality, let us
consider a scalar field $\Phi(x)$ which can interact with itself and
with other $n$ fields, say $\Psi_1(x)$, $\Psi_2(x)$,...$\Psi_n(x)$
according to some action \BE S[\Phi; \Psi_1, \Psi_2,...\Psi_n ] \EE
At this stage, both the nature of the $\Psi-$ fields (scalar,
fermion or gauge fields) and the structure of the action are
completely arbitrary. For instance, for $n=0$, the action could just
describe a single scalar field and be invariant under the simple
discrete reflection symmetry $\Phi \to -\Phi$. Or, for some $0 <k
\leq n$,  $\Psi_1$, $\Psi_2$,.. $\Psi_k$ could be other $k$ scalar
fields and the action be invariant for rotations in the
(k+1)-dimensional space ($\Phi$, $\Psi_1$, $\Psi_2$,.. $\Psi_k$).
Or, this type of global continuous symmetry could also be
transformed into a local symmetry provided gauge fields (with
corresponding gauge-fixing and gauge-compensating terms) were
introduced.

In this framework, the Green's functions of the $\Phi-$ field are
obtained from the generating functional in the presence of a source
$J(x)$ \BE \label{symbol0} Z[J]= \int
[d\Phi(x)][d\Psi_1(x)d\Psi_2(x)...d\Psi_n(x)] ~e^{ \int d^4x \Phi(x)
J(x) - S[\Phi; \Psi_1, \Psi_2,...\Psi_n ]} \EE Before exploring the
conditions for a non zero $\langle \Phi \rangle$, let us perform
formally the functional integration on the $\Psi-$ fields \BE
\label{symbol1} \int [d\Psi_1(x)d\Psi_2(x)...d\Psi_n(x)] ~e^{ -
S[\Phi; \Psi_1, \Psi_2,...\Psi_n ]} \equiv e^{ - S_{\rm eff}[\Phi]}
\EE so that the generating functional can be expressed as in a
one-component theory \BE \label{symbol2} Z[J]= \int [d\Phi(x)]~e^{
\int d^4x \Phi(x) J(x) - S_{\rm eff}[\Phi]} \EE

As mentioned above, the standard motivation to interpret the
broken-symmetry phase as a pure massive theory derives from the
chain relations (\ref{chain}). These are obtained by assuming
implicitly i) the regularity of $G(p=0)$ and ii) that the constant
background $\phi$ entering the expression for the full scalar field
$\Phi(x)$ \BE \Phi(x)= \phi + h(x) \EE can be kept `frozen' at one
of the two absolute minima $\pm v$ of a non convex effective
potential $V_{\rm NC}(\phi)$. To check the validity of these two
assumptions, let us first define the theory in some large 4-volume
$\Omega$. In this way, the possible space-time averages of the
field, namely \BE \phi={{1}\over{\Omega}}\int d^4x ~\Phi(x) \EE
represent the zero-momentum mode of the scalar field and, as such,
enter the full functional measure of the theory \BE \label{full}
\int[d\Phi(x)]...=\int^{+\infty}_{-\infty} d\phi\int[dh(x)]... \EE
In the above relation, the measure $[dh(x)]$ is over all quantum
modes with $p_\mu \neq 0$ that, for $\Omega \to \infty$, will
include arbitrarily small values of $|p|$. According to the standard
interpretation of the broken phase as a pure massive theory, the
zero-momentum limit of the connected propagator should be uniquely
determined, say \BE \lim_{p_\mu\to 0} G(p)={{1}\over{m^2_h}} \EE To
check this expectation, one can compute directly $G(p=0)$ from the
generating functional in the presence of a constant source $J$ and
then send $J \to 0$ and $\Omega \to \infty$. If there are no
subtleties associated with the infinite-volume limit, one should
obtain the same result.

By restricting to the case of a constant source $J$ in the
generating functional \BE \label{symbol} Z(J)=
\int^{+\infty}_{-\infty} d\phi~\exp (\Omega J \phi) \int[dh(x)]~
\exp - S_{\rm eff}[\phi + h] \EE  one can compute the field
expectation value \BE { { 1}\over{\Omega Z(J)}}~ {{d Z}\over{d
J}}=\langle \Phi \rangle_J\equiv \varphi(J) \EE and the
zero-momentum propagator \BE { { 1}\over{\Omega Z(J)}}~ {{d^2
Z}\over{d J^2}}=\Omega\varphi^2(J)+ {{d \varphi}\over{d J}} \EE To
express the connected parts, it is convenient to introduce the
generating functional for connected Green's function \BE W(J)=\Omega
w(J)= \ln {{Z(J)}\over {Z(0)}} \EE from which one obtains the field
expectation value \BE \label{jphi} \varphi={{d w}\over {d J}} \EE
and the zero-momentum connected propagator \BE G(p=0)={{d^2 w}\over
{d J^2}}={{d \varphi}\over{d J}} \EE In this framework, spontaneous
symmetry breaking corresponds to a non-zero value of $\varphi$ in
the double limit $J \to 0$ and $\Omega \to \infty$.

To study this limit, I will assume the standard condition for the
occurrence of spontaneous symmetry breaking, namely that the result
of the $h$-integration in Eq.(\ref{symbol}) can be expressed
formally, as in the loop expansion (see e.g.
\cite{maiani,ritschel}), in terms of some non-convex effective
potential which has absolute minima for $\phi \neq 0$ \BE
\int[dh(x)]~ \exp - S_{\rm eff}[\phi + h]= \exp-\Omega V_{\rm
NC}(\phi)\EE so that
 \BE Z(J)=\int^{+\infty}_{-\infty} d\phi~
\exp-\Omega(V_{\rm NC}(\phi)-J\phi) \EE In the $\Omega \to \infty$
limit, $Z(J)$ can then be evaluated in a saddle-point approximation
(see e.g. \cite{ritschel,rivers,brezin,zinnjustin}). In this
approximation, where the result is only expressed in terms of the
absolute minima  $\pm v$ of $V_{\rm NC}(\phi)$ and of its quadratic
shape there, say $V''_{\rm NC}(\pm v)=m^2_h$, the relevant relation
is \BE Z(J)\sim e^{\Omega J^2/(2m^2_h)} \cosh (\Omega J v) \EE up to
a $J-$independent proportionality factor. One thus obtains \BE
w(J)={{J^2}\over {2m^2_h}} + {{1}\over {\Omega}}\ln \cosh (\Omega J
v) \EE and \BE \label{invert} \varphi={{d w}\over {d J}}={{J}\over
{m^2_h}}+ v \tanh (\Omega J v) \EE \BE G(p=0)={{d^2 w}\over {d
J^2}}= {{1}\over {m^2_h}}+ {{\Omega v^2 }\over { \cosh^2(\Omega J v)
}}  \EE Since both $J$ and $\Omega$ are dimensionful quantities, it
is convenient to introduce dimensionless variables \BE
\epsilon={{J}\over {m^2_h v}} \EE and \BE y= \Omega m^2_h v^2 \EE so
that $\Omega J v = \epsilon y$. In this representation, one finds
\BE \label{phimean} \varphi= v (\epsilon + \tanh \epsilon y) \EE and
\BE G(p=0)={{1}\over {m^2_h}}\left[1 + y(1- \tanh^2 \epsilon
y)\right] \EE with the two limits $J \to \pm 0$ and $\Omega \to
\infty$ corresponding to $\epsilon \to \pm 0$ and $y \to \infty$.

It is clear that, when $\epsilon \to 0$, any non-zero limit of
$\varphi$ requires a non-zero limit of $\epsilon y$. Now, this limit
can be finite or infinite. If it is finite, say $\epsilon y \to X_0$
with $|X_0| < + \infty$, one finds $|\varphi|=v |\tanh X_0| < v$ and
an expectation value whose magnitude is reduced with respect to the
putative value $v$. In the same range, the zero-momentum propagator
$G(p=0)$ diverges for any value of $\varphi$. These results can be
rephrased by saying that, in the $\Omega \to \infty $ limit, both
$J=J(\varphi)$ and $G^{-1}(p=0)$ vanish in the open range $-v <
\varphi < +v$.

On the other hand, if the double limit $\epsilon \to \pm 0$ and $y
\to \infty$ corresponds to $\epsilon y \to \pm \infty$ one finds
$\varphi \to \pm v$. In this case, there are two different
possibilities for the zero-momentum propagator. Namely either
$G(p=0) \to 1/m^2_h$, if $ y(1- \tanh^2 \epsilon y) \to 0$, or
$G(p=0) \to \infty$, if $ y(1- \tanh^2 \epsilon y) \to \infty$.
Again this can be rephrased by saying that, in the $\Omega \to
\infty $ limit, if $J(\varphi)$ vanishes and $G(p=0)$ becomes a
double-valued function, then $\varphi$ tends to one of the absolute
minima of the underlying non-convex potential.

As anticipated, the above results admit a simple geometrical
interpretation in terms of the Legendre-transformed, quantum
effective potential $V_{\rm LT}(\varphi)$. This is defined through
the relation \BE V_{\rm LT}(\varphi)= J \varphi - w(J) \EE after
inverting $J=J(\varphi)$ so that \BE {{d V_{\rm LT}(\varphi)}\over
{d \varphi}}= J(\varphi)\EE and \BE {{d^2 V_{\rm LT}(\varphi)}\over
{d \varphi^2}}= {{d J}\over {d \varphi}}= G^{-1}(p=0) \EE Therefore,
by using Eq.(\ref{invert}) to define $J(\varphi)$, one obtains \BE
\label{pro2} G^{-1}(p=0)={{m^2_h }\over {1 + y (1- \tanh^2 \epsilon
y) }} \EE Then, the divergent value of $G(p=0)$, found in connection
with all values $|\varphi| < v$, corresponds to the inner region
where the Legendre-transformed effective potential becomes exactly
flat in the infinite-volume limit.

Analogously, the two possible solutions for $G(p=0)$, when $\epsilon
y \to \pm \infty$ and $\varphi \to \pm v$, correspond to compute
left and right second derivatives of $V_{\rm LT}(\varphi)$ at
$\varphi = \pm v$. To this end, let us compute the right second
derivative for $\varphi=+v$, i.e. the value of $G^{-1}(p=0)$ when
$\varphi \to v^+$, and set $\varphi= v + |\Delta \varphi|
> v$. From Eq.(\ref{phimean}),
one has $\epsilon \sim {{ |\Delta \varphi|}\over{ v}}$ up to terms
${\cal O}( e^{-\epsilon y})$ that vanish exponentially when $y \to
\infty$ and $\epsilon y \to \infty$. Thus, one finds
 \BE y (1- \tanh^2 \epsilon y) \to  0 \EE
and \BE G^{-1}(p=0)\to m^2_h\EE (analogous results hold for the left
derivative at $\varphi=-v$, i.e. for $\varphi= -v -|\Delta \varphi|
<- v$).

Let us now consider the left second derivative for $\varphi=v$, i.e.
the value of $G^{-1}(p=0)$ when $\varphi \to v^-$, and set $\varphi=
v - |\Delta \varphi| < v$. Here, the situation is different since,
now, one cannot set $\epsilon \sim {{ -|\Delta \varphi|}\over{ v}}$
to solve Eq.(\ref{phimean}). In fact, $ \epsilon y$ has to be large
and positive in order $\tanh \epsilon y$ to be a positive number
slightly smaller than unity. Rather, one can use Eq.(\ref{phimean})
to replace \BE \tanh\epsilon y= 1-\epsilon-{{|\Delta
\varphi|}\over{v}} \EE and show that \BE y [1-\tanh^2\epsilon y] =
y\left[ 2\epsilon + 2{{|\Delta \varphi|}\over{v}} + {\cal O}(
\epsilon^2)\right] > 2\epsilon y \to \infty \EE Then the left second
derivative vanishes \BE \label{pro3}
 G^{-1}(p=0)\sim  {{m^2_h}\over{ 1 + 2\epsilon y }} \to 0
\EE (analogous results hold for the right derivative at
$\varphi=-v$, i.e. when $\varphi= -v +|\Delta \varphi|
> - v$). In conclusion, at the absolute minima $\pm v$ of the
non-convex potential, $G^{-1}(p=0)$ becomes a two-valued function in
the infinite-volume limit of the theory.

Notice that, independently of any specific calculation, the gapless
solution $G^{-1}(p=0)=0$ is also needed for a consistent
interpretation of symmetry breaking with a convex-downward quantum
effective potential. In fact, if there were only massive excitations
in the spectrum one expects a non-degenerate ground state and,
therefore, an effective potential with only one minimum. Due to the
underlying reflection symmetry of the theory, this unique minimum
could only be $\varphi=0$. On the other hand, if there were a
gap-less branch in the spectrum, by adding a sufficiently large
number of these excitations in the zero 3-momentum mode, one could
construct new translational invariant states with different values
of $\varphi$ and the same energy. This construction could be done by
respecting the underlying $\varphi \to -\varphi$ symmetry, at least
in some range of $|\varphi|$, and leads to the type of degenerate
ground state associated with a flat effective potential. As
discussed in the following section, it is precisely this type of
degeneracy that is responsible for the existence of infinitesimally
weak long-range forces.

\section{An infinitesimal $1/r$ long-range potential}

The existence of two solutions for $G^{-1}(p=0)$, as deduced in
Sect.3, is just a consequence of spontaneous symmetry breaking in
the infinite-volume limit of the theory. As such, it does not imply
any specific functional form of $G(p)$. However, by exploiting the
analogy with superfluid $^4$He, one expects the full $G(p)$ to
correspond to a suitable interpolation between gap-less and massive
solutions. It is interesting that such interpolation can also be
deduced \cite{plb09} by using the strong constraints on the possible
structure of $G(p)$ placed by the accepted `triviality'
\cite{triviality} of the scalar self-interacting theories in four
space-time dimensions.

In fact, for the continuum theory, defined in the limit of infinite
ultraviolet cutoff, `triviality' dictates a gaussian set of Green's
functions, no observable dynamics at any value of the 4-momentum
$p_\mu \neq 0$ and the standard free-field type form
$G^{-1}(p)=(p^2+m^2_h)$. Still, consistently with these constraints,
one cannot exclude a discontinuity of the truncated Green's
functions in the zero-measure, Lorentz-invariant subset $p_\mu=0$.
This plays a fundamental role in translational invariant vacua
characterized by space-time constant expectation values of local
operators such as $\langle \Phi \rangle$. Therefore, by accepting
`triviality', the only possible not-entirely-trivial continuum limit
of the connected propagator has $G^{-1}(p)= (p^2+m^2_h)$ for any
$p_\mu \neq 0$ with the exception of a discontinuity at $p_\mu=0$
where $G^{-1}(p=0)=0$.

Let us now consider the finite-cutoff theory. Here, in the presence
of an ultraviolet cutoff $\Lambda$, the distinction between
$p_\mu=0$ and $p_\mu \neq 0$ has no obvious meaning. One can always
consider a whole set of `infinitesimal' (but non-zero) momentum
values, such as $|p|\sim m^2_h/\Lambda$, $|p|\sim
m^3_h/\Lambda^2$,... that however all approach the same $p_\mu=0$
value when $\Lambda\to \infty$. Therefore, in the cutoff theory, if
one wants to obtain a continuum limit where $G^{-1}(p=0)=0$, at a
certain point, i.e. for sufficiently small momenta, one should
necessarily replace the standard massive form $G^{-1}(p)\sim (p^2 +
m^2_h) \to m^2_h$ with some different behaviour for which $G^{-1}(p)
\to 0 $. For this reason, the sharp singularity of the continuum
theory will be replaced by a smooth behaviour in the cutoff theory.
Then, even though the continuum theory has only massive, free-field
excitations, the cutoff version would exhibit non-trivial
qualitative differences, as weak long-range forces, that cannot be
considered uninteresting perturbative corrections.

For a quantitative description, one can write the connected
propagator of the cutoff theory in a general interpolating form, say
\cite{plb09}\BE \label{inter}G^{-1}(p)=(p^2 +m^2_h) f(p^2/\delta^2)
\EE In order to reproduce the mentioned continuum-limit behaviour,
the function $f(p^2/\delta^2)$ has to refer to some infrared
momentum scale $\delta\neq 0$ (with $\delta/m_h\to 0$ when
$m_h/\Lambda \to 0$, as for $\delta\sim m^2_h/\Lambda$) in such a
way that \BE \lim_{\delta\to 0}f(p^2/\delta^2)=1~~~~~~~~~~~~~(p_\mu
\neq 0)\EE with the only exception \BE \lim_{p_\mu\to
0}f(p^2/\delta^2)=0~~~~~~~~~~~~~\EE (think for instance of
$f(x)=\tanh(x)$, $f(x)=1 - \exp(-x)$, $f(x)=x/(1+x)$,...).

As anticipated, to understand what kind of long-range effects
 in coordinate space are associated with such propagator
for the scalar field, one has to consider the standard Fourier
transform of the zero-energy propagator $G({\bf p},p_4=0)$ \BE
\label{fourier} D({\bf r})= \int {{d^3{\bf p}}\over{(2\pi)^3}}
 {{e^{i {\bf p}\cdot{\bf r}}}\over{ ({\bf p}^2 + m^2_h) f({\bf p}^2/\delta^2) }}\EE
that in the case of a free massless scalar field, $G({\bf
p},p_4=0)=1/{\bf p}^2$, gives a $1/r$ potential.

Now, a straightforward replacement $ f({\bf p}^2/\delta^2) =1$ would
produce the Yukawa potential $e^{-m_h r}/r$. However, if we consider
the finite-cutoff theory, we have to take into account the region $
{\bf p}^2 \ll \delta^2$ where the relevant limiting relation is
rather \BE \lim_{{\bf p} \to 0}f({\bf
p}^2/\delta^2)=0~~~~~~~~~~~~~\EE For this reason, since the dominant
contribution for $r \to \infty$ comes from ${\bf p}=0$, where the
denominator in (\ref{fourier}) vanishes, there will be long-range
forces. In this case, by expanding around ${\bf p}=0$ and using the
Riemann-Lebesgue theorem on Fourier transforms \cite{plb09},
whatever the detailed form of $f(x)$ at intermediate $x$, the
leading contribution at asymptotically large $r$ will be 1/r . One
thus gets \BE \label{asy}\lim_{r \to \infty}~D({\bf r})=
D_{\infty}({\bf r})={{ \delta^2}\over{f'(0)m^2_h}}~{{1}\over{4\pi
r}}\EE all dependence on the interpolating function being contained
in the factor $f'(0)$ expected to be ${\cal O}(1)$. In this way, the
same asymptotic $1/r$ trend as in Ref.\cite{ferrer} is obtained by
using only general properties of the underlying quantum field
theory.

To put some numbers (in units $\hbar=c=1$), let us consider for
definiteness the scenario $\delta\sim m^2_h/\Lambda$ which is
motivated by the relations of Sect.2, namely $\Lambda \sim 1/a$,
$m^2_h \sim {n}a$ and the hydrodynamic-limit relation $\delta\sim
1/r_{\rm mfp}\sim {n}a^2 $. By fixing the same values of Sect.2,
namely $a\sim 10^{-33}$ cm, $1/m_h=\xi_h \sim 10^{-17}$ cm, and
$r_{\rm mfp} \sim 10^{-1}$ cm, let us consider the couplings of the
singlet standard model Higgs boson. Two fermions $i$ and $j$, of
masses $m_i$ and $m_j$, couple to it with strength $y_i=m_i/v$ and
$y_j=m_j/v$ and thus feel the instantaneous potential \BE V({\bf r})
= -y_i y_j D({\bf r}) \EE From the previous analysis, besides the
short-distance Yukawa potential governed by the Fermi constant
$G_F\equiv 1/v^2$ \BE \label{fourier5} V_{\rm yukawa}({\bf r}) =
-{{G_F m_i m_j}\over{4\pi r}}e^{-m_h r}\EE (that dominates for $r
\lesssim 1/m_h$), they would feel the asymptotic potential
associated with Eq.(\ref{asy}). This can be conveniently expressed
as \BE \label{fourier6} \lim_{r \to \infty}~V({\bf
r})=V_{\infty}({\bf r}) = -{{G_{\infty} m_i m_j}\over{ r}} \EE with
the effective coupling  \BE \label{infty} G_{\infty}=
{{\delta^2}\over{4\pi f'(0)m^2_h }}~G_F \sim
{{10^{-33}G_F}\over{f'(0)}} \EE Strictly speaking, this asymptotic
potential represents a `cutoff artifact' since the continuum theory
has only massive, free-field excitations, with the only exception of
a discontinuity at $p_\mu=0$ where $G^{-1}(p)=0$. At least, this is
the only possible remnant of symmetry breaking allowed by exact
Lorentz invariance and `triviality'. However in the cutoff theory,
where one expects a smooth behaviour, the deviation from the massive
form will necessarily extend, from the zero-measure set $p_\mu=0$,
to an infinitesimal momentum region $\delta$. It is this
infinitesimal momentum region, where the propagator will look like
in a massless theory, to produce the long-range $1/r$ potential of
infinitesimal strength $\delta^2/m^2_h$.

\section{The scalar coupling in the hydrodynamic limit}

As discussed at the end of Sect.2, the infrared limit $p_\mu \to 0$
of the propagator determines the long-range scalar fluctuations that
have been denoted as the $s-$ field. In terms of this field, one can
re-formulate the two-body interaction \BE \label{fourier7}
V_{12}({\bf r}_1 -{\bf r}_2) = -{{G_{\infty} M_1 M_2}\over{ |{\bf
r}_1 -{\bf r}_2|}} \EE as \BE \label{fourier8} V_{12}({\bf r}_1
-{\bf r}_2) =M_1s({\bf r}_1) \EE with \BE s({\bf r}_1)=
-{{G_{\infty} M_2}\over{ |{\bf r}_1 -{\bf r}_2|}}\EE or as \BE
\label{fourier9} V_{12}({\bf r}_1 -{\bf r}_2) =M_2s({\bf r}_2) \EE
with \BE s({\bf r}_2)= -{{G_{\infty} M_1}\over{ |{\bf r}_1 -{\bf
r}_2|}}\EE In this way, provided the Newton constant $G_N$ is
identified with the infinitesimal coupling $G_\infty$, one gets the
idea of the Newtonian potential as a fundamental excitation of the
scalar condensate. The two equivalent ways to write down the same
two-body interaction, namely as the particle 1 in the field
generated by the particle 2 or as the particle 2 in the field of the
particle 1, would then express the identity of `active' and
`passive' gravitational masses and their equality to the inertial
mass generated by the vacuum structure.

For a $N-$ body system, one can use the relation \BE \Delta~(
{{1}\over{r}})= -4\pi \dl \EE that expresses the asymptotic
propagator as the Green's function of the Poisson equation with a
given mass density, i.e. \BE \Delta s(x)= 4\pi G_N \sum M_n
\delta^{(3)}({\bf r}- {\bf r}_n) \EE Notice that, by writing down
such Poisson equation, one is assuming that by adding more and more
sources the resulting $s-$fields superpose linearly. This is only
true if the residual self-interaction effects of the $s-$field are
negligible. Formally, in the broken-symmetry phase of a $\Phi^4$
theory, these self-interaction effects start to ${\cal O}(s^3)$ and
might show up as ${\cal O}(s^2)$ in the equations of motion. The
fact that these terms can be neglected (in ordinary circumstances)
is consistent with the idea that $s(x)$ represents an excitation of
the vacuum in a `trivial' theory, where physical states should
exhibit no observable self-interaction.

Thus,  the derivation of the Newtonian potential from the vacuum
structure could be considered complete for all elementary particles
that couple directly to the fundamental Higgs field. In the standard
model, these elementary particles are the leptons and quarks and,
for them, the coupling to any scalar field can simply be considered
a local re-definition of their mass parameters. To this end, one
should start from the basic elementary yukawa couplings \BE
\label{yu1} {\cal L}_{\rm yukawa}= -y_f \bar{\psi}_f\psi_f \Phi \EE
and decompose the full scalar field $\Phi(x)$ in the sum of its
vacuum expectation value $\langle \Phi \rangle=v$, of the long-range
component $v s(x)$ (for wave vectors ${\bf k} \lesssim \delta$) and
of the short-range part (for ${\bf k}
> \delta$). In this way, by defining $m_f=y_fv$,
(and dropping the short-wavelength components) one gets trivially
\BE \label{yu2} {\cal L}_{\rm yukawa}= -m_f \bar{\psi}_f\psi_f (1
+s) \EE so that a non-zero $s$ amounts to re-scale all elementary
fermion masses according to $m_f \to m_f (1+s)$.

A possible objection can arise when considering the hadronic states,
as nucleons, nuclei,.. that, differently from leptons and quarks,
have no tree-level coupling to the scalar field as in
Eq.(\ref{yu1}). Besides depending on the quark masses, their masses
refer to other basic parameters of the vacuum, namely the gluon and
the chiral condensate. In this case, why should one introduce the
same overall type of mass re-scaling ?

To see this, let us first consider the unperturbed situation where
$s=0$. In this limit, the mutual interactions among the various
scalar, gluon and quark condensates give rise to suitable relations
arising from the minimization of the overall energy density. One can
express these relations as \BE
        \alpha_{\rm QCD}\langle F^a_{\mu\nu}F^a_{\mu \nu}\rangle_o= c_1 v^4
\EE and \BE
            m^{(o)}_q \langle \bar{\psi}_q \psi_q \rangle_o= c_2 v^4
\EE where $c_1$ and $c_2$ are dimensionless numbers and
$\langle...\rangle_o$ denotes the unperturbed vacuum expectation
values for $s=0$.

Now, let us consider an external perturbation that induces
long-wavelength oscillations of the scalar condensate so that $v \to
v(1+ s)$. Let us also assume that QCD has only short-range
fluctuations whose wavelengths are much smaller than $1/\delta$. In
this situation, $v(1+ s)$ can be considered to define a new local
vacuum, whose variations occur over regions that are much larger
than the QCD scale. To this new scalar field the quark masses and
all relevant expectation values of the gluon and quark operators
will (`adiabatically') be adjusted according to the new minimization
relations \BE
        \alpha_{\rm QCD}\langle F^a_{\mu\nu}F^a_{\mu \nu}\rangle= c_1
v^4(1+s)^4 \EE and \BE
            m_q \langle \bar{\psi}_q \psi_q \rangle=
c_2 v^4(1+s)^4 \EE In this way, the overall result of a non-zero $s$
is equivalent to a rescaling of the QCD scale parameter \BE
\Lambda_{\rm QCD} \to \Lambda_{\rm QCD}(1+ s) \EE of the chiral
condensate \BE \langle \bar {\psi_q} \psi_q \rangle \to \langle \bar
{\psi_q} \psi_q \rangle (1+ s)^3 \EE and of the quark masses \BE
               m_q \to m_q (1+ s)
\EE Therefore, any combination with the dimension of a mass, say
$\Lambda^4_{\rm QCD}/\langle \bar {\psi_q} \psi_q \rangle$ or
$\langle \bar {\psi_q} \psi_q \rangle/\Lambda^2_{\rm QCD}$, that is
expressed in terms of the elementary quark masses as well as of the
expectation value of both quarks and gluon condensates, will also
undergo the same rescaling $ (1+ s)$. This holds for the nucleon
mass $m_N$ \BE
               m_N \to m_N (1+ s)
\EE for the mass of the nuclei and, more in general, for all mass
parameters and binding energies that can be expressed in terms of
the basic quantities of the theory, namely the masses of the
elementary fermions and the vacuum condensates. This is due to the
extremely large $s-$field wavelengths as compared to any elementary
particle, or nuclear, or atomic scale. Thus $s(x)$ couples
universally to the various forms of matter and one obtains the same
overall type of re-scaling Eq.(\ref{rescaling2}) foreseen in the
Introduction.

Finally, let us consider the extension to the case of variable
fields. To this end, one should take into account both the full
$p^2-$dependence in the propagator and replace, in the basic
coupling to the scalar field, the mass density with the trace of the
energy momentum tensor. This replacement can be understood by
considering a fermion field $\psi_f$ that describes a sharply
localized wave packet with momentum ${\bf{p}}$ and velocity ${\bf
v}$, i.e. such that \BE\int d^3 x \langle \bar{\psi}_f\psi_f\rangle=
{{m_f}\over{\sqrt{ {\bf{p}}^2 + m^2_f }}}=\sqrt{1 - {\bf{v}}^2}\EE
If $s(x)$ does not vary appreciably over the localization region,
one gets from Eq.(\ref{yu2}) the classical action \BE \label{cla}
             \int d^4 x{\cal L}_{\rm yukawa}
            =
- m_f \int d\tau(1+ s(x)) \EE for a point-like particle interacting
with a scalar field $s(x)$. In the last expression
$x_\mu=x_\mu(\tau)$ and $d\tau=dt \sqrt{1 - {\bf{v}}^2}$ denotes the
proper-time element of the particle.

As for any linear coupling, this relation, that gives the action of
a particle in an external $s-$field, can also be used to express the
$s-$field that is generated by a given source. In this case, by
using Eq.(\ref{inter}) to describe the full $p^2-$dependence of the
propagator for $p_\mu \to 0$, taking the Fourier transform and
identifying $G_\infty=G_N$, one obtains an equation of motion that
is valid for large space-time separation from the sources. For a
many-particle system, this amounts to \BE \square s(x)= 4\pi G_N
T^\mu_\mu(x) \EE where \BE \label{tmumu}
         T^{\mu}_{\mu}(x) \equiv \sum_n
M_n \sqrt{1 - {\bf{v}}^2_n}~ \delta^3 ( {\bf{x}} - {\bf{x}}_n(t) )
\EE Notice that the trace of the energy-momentum tensor can be
considered the Lorentz-invariant density of inertia. In fact, as
discussed by Dicke \cite{varenna}, when averaged over sufficiently
long times (e.g. with respect to the atomic times), by the virial
theorem \cite{champ}, the spatial integral of $T^\mu_\mu$ represents
the total energy of a bound system,  i.e. includes the binding
energy. Therefore, for microscopic systems whose components have
large ${\bf{v}}^2/c^2$ but very short periods, this definition
becomes equivalent to the rest energy. On the other hand, for
macroscopic systems, that have long periods but small
${\bf{v}}^2/c^2$, the definition gets close to the mass density
\footnote{In lowest order, the source of the $s-$field is just the
trace of the energy-momentum tensor. However, the effective metric
(\ref{basic02}) is not obtained from the flat-space metric through
an overall conformal factor $e^{2s}\sim 1+2s +...$. Thus, there is a
basic difference with pure scalar theories of gravitation as, for
instance, Nordstr\"om gravity \cite{nordstrom}.}.

\section{Summary and conclusions}

Several authors (see e.g. the review in \cite{barcelo}) have pointed
out interesting analogies between what we call `Einstein gravity'
and the hydrodynamic limit of many condensed matter systems in flat
space. Although one does not expect to reproduce exactly the same
features of classical General Relativity, still there is some value
in exploring this type of correspondence. In fact, beyond the simple
level of an analogy, there might be a deeper significance whenever
the properties of the hypothetical underlying medium could be
matched with those of the physical vacuum of present particle
physics. In this case, the vacuum condensates (Higgs, gluon,
chiral,...) of electroweak and strong interactions, that play a
crucial role for fundamental phenomena, as particle mass generation
and quark confinement, could also represent a `bridge' from particle
physics to gravity.

To explore this possibility, one could start on a general ground by
first representing the physical flat-space vacuum as a medium in
which there is a scalar fluctuation field $s(x)$ that, on a
coarse-grained, acts as in Eq.(\ref{rescaling1}). As discussed in
the Introduction, this leads naturally to the isotropic form of the
effective metric \BE \label{basic05}
         {g}_{\mu \nu}[s] \equiv
{\rm diag}({{1
}\over{\lambda^2(s)}},-\lambda^2(s),-\lambda^2(s),-\lambda^2(s)) \EE
so that, by expanding in powers of $s(x)$ with the natural condition
$\lambda(0)=1$ and choosing the units so that \BE
\label{rescaling22} {{1}\over{\lambda(s)}}= 1+s +{\cal O}(s^2) \EE
one arrives to the weak-field line element \BE \label{basic03}
         {g}_{\mu \nu}[s]\equiv
{\rm diag}((1+2 s) ,-(1-2s),-(1-2s),-(1-2s)) \EE Then, by comparing
with experiments in weak gravitational field, one gets a
coarse-grained identification of the $s-$field with the Newtonian
potential, i.e. \BE \label{ident} s \sim U_N= - G_N \sum_i
{{M_i}\over{|{\bf r} - {\bf r}_i|}} \EE and it becomes natural to
ask whether the ultimate origin of such scalar field $s\sim U_N$
could be found in the presently accepted vacuum condensates of
particle physics \footnote{As mentioned in the Introduction, the
idea that Bose condensates can provide various types of
gravitational dynamics is not new. In particular, one is not bound
to a long-range $1/r$ behaviour and can also consider the
possibility of short-range analogues of the Newtonian potential
\cite{sindoni}.}. While $1/r$ long-range potentials are indeed
expected among bodies placed in a Bose condensate of spinless quanta
\cite{ferrer}, one should try to understand from first principles
why the magnitude of the relevant $1/r$ potential is so small in
units of the physical coupling strength set by the Fermi constant
$G_F$. To this end, one can use i) the basic two-valued nature of
the zero-momentum connected scalar propagator in the broken-symmetry
phase and ii) the `triviality' of contact, point-like interactions
in 3+1 dimensions. These two requirements, implying that the
far-infrared, gap-less region of the spectrum of the broken phase
has to become of zero-measure in the continuum limit of the theory,
could naturally explain the origin of a hydrodynamic coupling
$G_\infty$ that is infinitesimally weak in units of $G_F$. If this
coupling is identified with the Newton constant $G_N$, one has a toy
model of gravity, i.e. restricted to a definite class of phenomena,
that could be used as a first approximation to reduce the present
conceptual gap with particle physics.

At the present, the approach has some interesting aspects but also
several limitations. The interesting features consist in a
potentially simple explanation of the hierarchical pattern of scales
observed in Nature. By describing the scalar condensate as a true
physical medium, made up of elementary constituents, one discovers
that the broken phase is much richer than what is usually believed.
In fact, by approaching the continuum limit of the underlying scalar
quantum field theory, one could easily handle a hierarchy of scales
containing an extremely small scattering length $a\sim 10^{-33}$ cm,
a typical elementary-particle scale $ 1/\sqrt {{n}a}\sim 10^{-17}$
cm and a macroscopic length $1/{{n}a^2}\sim 10^{-1}$ cm that can be
used to mark the onset of a hydrodynamic regime.

At the same time, one can easily understand basic properties of the
gravitational interaction. For instance, the exact equality of both
passive gravitational mass and active gravitational mass with the
same inertial mass generated by the vacuum structure is
straightforward in this picture but is not so obvious in a pure
geometrical formalism where inertial and gravitational masses could
differ by an universal, but otherwise arbitrary, proportionality
constant.

Finally, as in other frameworks based on an underlying superfluid
vacuum \cite{volo}, the relevant curvature effects will be orders of
magnitude smaller than those expected by solving Einstein's
equations with the full energy-momentum tensor as a source term. In
fact, by definition, the large condensation energy of the
unperturbed vacuum does not play any observable role.

On the other hand, there are also several limitations. For instance
the numerical relations have still a certain degree of uncertainty
and I have been unable to determine exactly the value of the
infinitesimal coupling $G_\infty$ in Eq.(\ref{infty}). Even knowing
the Higgs mass parameter $m_h$, the interpolating function in the
propagator (and thus the proportionality factor $f'(0)$) remains
unknown since, to this end, one should solve a non-linear integral
equation. Analogous problems arise if one wants to fix the precise
size of the infrared momentum scale $\delta$ in terms of the mass
parameter $m_h$ and of the cutoff $\Lambda$. Also in this case, it
is not easy to improve on the simple order of magnitude estimate
$\delta \sim m^2_h/\Lambda$ that is suggested by the identification
of the length scale $1/\delta$ with the mean free path of the
elementary constituents. To determine the precise values of the
various parameters, the shape of the potential in the intermediate
region and check the overall consistency of the picture with the
precise measurements of gravity, at and below the millimeter scale
\cite{eotwash}, a new generation of lattice calculations would be
important. As discussed in Ref.\cite{plb09}, in fact, precise
measurements on large lattices of the slope of the scalar propagator
in the infrared region could be used to extrapolate the trend to
both the continuum and infinite-volume limit of the theory.

Another problem concerns the structure of the neglected ${\cal
O}(s^2)$ effects in $\lambda(s)$. These arise from higher-order
Feynman graphs where more and more scalar quanta are exchanged
between two sources and interact with each other. In principle, one
could try a direct re-summation of these higher-order, infrared
effects analogously to the Bloch-Nordsieck exponentiation method in
QED. However, re-summing infrared effects in self-interacting
theories is not so simple and it will not be easy to find the answer
in this way. At the same time, non-leading $1/r^2$ effects are also
expected from the higher-derivative expansion around $p=0$ of the
propagator entering the same one-boson exchange graph.

Now, as mentioned in the Introduction, in an emergent-gravity
approach, it is natural to consider the hydrodynamic limit of the
underlying superfluid vacuum (embodied in the peculiar aspects of
the Newtonian potential) as the crucial ingredient that induces
forms of physical equivalence requiring a general-covariant
formulation. Therefore, the structure of these higher-order terms in
the effective metric should correspond to a solution of Einstein's
field equations. However, the metric structure depends on the
energy-momentum tensor that, in general, might depend on $s(x)$. To
better appreciate this point, let us first consider Einstein's field
equations for the weak-field metric when $s \sim U_N$ \BE
\label{basic07}
         {g}_{\mu \nu}[U_N]\equiv
{\rm diag}((1+2 U_N) ,-(1-2U_N),-(1-2U_N),-(1-2U_N)) \EE As
discussed by Synge \cite{synge}, this corresponds to bodies with
energy density \BE T^{\rm matter}_{00}\sim \Delta U_N =\sum_i M_i
\delta^{(3)}({\bf r}- {\bf r}_i ) \EE embedded in a medium with
energy density \BE T^{\rm medium}_{00}\sim (\nabla U_N)^2 \EE As
mentioned in the Introduction, beyond this lowest-order
approximation, exact solutions have been obtained for the two cases
of the Schwarzschild and Yilmaz metrics. They are both in agreement
with the weak-field tests but differ non-trivially in the
strong-field limit. The physical reason for this difference consists
in the presence of an energy density ${\cal E}$ associated with the
scalar field itself, in particular in the voids among the elementary
constituents that exist inside massive bodies whereas the
Schwarzschild metric corresponds to the limit ${\cal E}=0$. Thus, it
is understandable that, with different forms of such energy density,
there could be an effective screening of the Schwarzschild
singularity expected for a single point-like mass.

This could have non-trivial phenomenological implications for
astrophysics and even for cosmology. For instance, in the case of
the Yilmaz metric, for a given equation of state for dense matter,
stable stellar objects of larger mass might be allowed
\cite{robertson} \footnote{This resembles somehow the increase in
the critical charge of super-heavy nuclei, namely $Z_c\sim$173
rather than $Z_c=$137, that is obtained \cite{greiner} by replacing
the point-like charge approximation with an extended charge
distribution.}. Analogously, since in the Yilmaz metric there is no
limit to the gravitational red-shift of light emitted by dense
matter, one could find alternative explanations for the
controversial huge quasar red-shifts, a large part of which could be
interpreted as being of gravitational (rather than cosmological)
origin \cite{clapp}.

\vskip 10 pt

\centerline{{\bf Acknowledgments}} \vskip 5 pt I thank Prof. Attilio
Agodi for many useful discussions.

\end{document}